\documentclass[
 aip,
 amsmath,
 amssymb,
 reprint,
 jcp,
]{revtex4-1}

\usepackage[utf8]{inputenc}
\usepackage[T1]{fontenc}
\usepackage{mathptmx}
\usepackage{etoolbox}

\usepackage{graphicx}
\usepackage{dcolumn}
\usepackage{bm}
\usepackage{hyperref}
\usepackage{gensymb}

\usepackage[shortlabels]{enumitem}

\date{\today}

\begin{document}

\title{Accelerating structure search using atomistic graph-based classifiers}

\author{Andreas Møller Slavensky}
\author{Bjørk Hammer}
    \email{hammer@phys.au.dk}
\affiliation{Center for Interstellar Catalysis, Department of Physics and Astronomy, Aarhus University, Aarhus C, DK‐8000 Denmark}

\begin{abstract}
We introduce an atomistic classifier based on a combination of spectral graph theory and a Voronoi tessellation method. 
This classifier allows for the discrimination between structures from different minima of a potential energy surface, making it a useful
tool for sorting through large datasets of atomic systems. We incorporate the classifier as a filtering method in the 
Global Optimization with First-principles Energy Expressions (GOFEE) algorithm. Here it is used to filter out structures from exploited
regions of the potential energy landscape, whereby the risk
of stagnation during the searches is lowered. We demonstrate the usefulness
of the classifier by solving the global optimization problem of 2-dimensional pyroxene, 3-dimensional olivine, Au$_{12}$, and Lennard-Jones LJ$_{55}$ and LJ$_{75}$ nanoparticles.
\end{abstract}

\maketitle

\section{Introduction}

In recent years, machine learning (ML) models have become increasingly relevant in materials science. They can be used to predict various properties
of different atomistic compositions at the level of density functional theory (DFT) or above, but at a fraction of the computational cost,
enabling for large-scale simulations that would not be possible using first-principle methods.
They have been used to aid global optimization (GO) algorithms \cite{huang_material_2017,deringer_data-driven_2018,paleico_global_2020,arrigoni_evolutionary_2021,kaappa_beacon_2021,han_unfolding_2022,pierrelourenco_automatic_2022,jung_machine-learning_2022,merte_2022,chen_automated_2022,local_model,larsen_machine-learning-enabled_2023,shi_accessing_2023,kim_recalibrating_2024},
perform molecular dynamics simulations \cite{li_molecular_2015,deringer_machine_2017,jindal_spherical_2017,roongcharoen_oxidation_2023,xie_uncertainty-aware_2023},
calculate phase diagrams \cite{jinnouchi_phase_2019,rosenbrock_machine-learned_2021,rossignol_machine-learning-assisted_2024}, and predict infrared spectra \cite{gastegger_machine_2017,beckmann_infrared_2022,tang_machine_2023}.  
Large datasets of structural configurations and calculated properties are required in order to train reliable ML models.
These can be generated on-the-fly using active learning methods\cite{ouyang_global_2015,smith_less_2018,sivaraman_machine-learned_2020,vandermause_--fly_2020,timmermann_data-efficient_2021,yang_machine-learning_2021,van_der_oord_hyperactive_2023,wang_magus_2023}, or collected from the resulting structures
from GO searches, but these tasks can be quite cumbersome and time consuming. 
Recently, several databases have been constructed and made publicly available\cite{irwin_zinc_2005,ruddigkeit_enumeration_2012,jain_commentary_2013,winther_catalysis-huborg_2019,schreiner_transition1x_2022}, potentially reducing the time required
to obtain stable and well-behaving ML models. 

When dealing with large datasets, a reliable classifier can be useful in order to distinguish structures from each other.
This can be relevant to ensure large diversity in a training set for a ML model in order to avoid overfitting.
For example, after a GO search\cite{andersen_descriptors_2023}, a lot of structures from the same energy basin are often found,
and it would be convenient to filter out a subset of these structures from such a collection without decreasing its quality.

To construct such a classifier, a computational method for encoding and describing a given structure is needed. Through these descriptions,
it should be possible to quantify whether two structures are identical or not. In the context of ML, there exist different methods to encode structures in order to obtain a model which 
is invariant to certain transformations, i.e., translations and rotations. Examples of different categories
of descriptors are global feature descriptors\cite{Fingerprint2010,schutt_how_2014,hansen_machine_2015}, where a single representation is constructed for the entire structure, and local feature descriptors\cite{Behler2011,SOAP2013,drautz_atomic_2019},
where each atom in a configuration is given its own representation. Recently, graph representations\cite{rupp_fast_2012,faber_crystal_2015} have become a popular way to represent
structures, and graph neural networks produce state-of-the-art results in many ML tasks.\cite{zhu_fingerprint_2016,schutt_schnet_2018,schutt_equivariant_2021,batzner_e3-equivariant_2022,batatia_mace_2022,xu_2022,deng_chgnet_2023} From these representations,
a measure of similarity between configurations can be constructed, which can be used to ensure a large diversity in
the training set for the ML models\cite{bartok_machine_2017,tahmasbi_large-scale_2021,lee_staged_2023}.

Another case for having a strong classifier is to guide certain GO algorithms. Some algorithms, such as evolutionary algorithms (EA) \cite{hartke_global_1995,EA2003,vilhelmsen_genetic_2014},
rely on sampling a subset of previously evaluated structures to create a population, in this paper referred to as a $sample$. From these structures, new configurations are created to advance the search.
Relying on a sample of low-energy structures can potentially lead to stagnation in the search, since the sampled structures might
not contain the required configurational diversity that is needed to explore the potential energy surface (PES) and find the global energy minimum (GM) structure.
In such cases, a classifier can be used to quantify the usefulness of certain structures and eventually filter unwanted configurations
out of the sample. Similar ideas have previously been used to improve other GO algortihms, such as in EA\cite{hartke_global_1999}, where the sample is kept diverse through distance criteria,
and in minima hopping\cite{goedecker_minima_2004,krummenacher_performing_2024}, where the molecular dynamics temperature is gradually increased when the search is stuck in an energy funnel.
 This can increase the explorative nature of the search, hopefully leading to a more robust algorithm.

In this work, we propose a classifier that is based on spectral graph theory\cite{wilson_study_2008,wills_metrics_2020}, which is used to generate simple graph representations of structures. 
We construct a modified adjacency matrix, which describes the existence of bonds between atoms within a given structure. We combine this with a 
Voronoi tessellation method\cite{okeeffe_proposed_1979} to improve its stability towards subtle changes to atomic positions. The Voronoi tessellation method has previously been 
used to describe coordination numbers in crystals \cite{pan_benchmarking_2021}, where it was shown to be robust to small changes in the atomic positions, and to extract features for crystal structure representations\cite{ward_including_2017,park_developing_2020}.
This classifier is discontinuous, enabling us to easily distinguish between structures
by directly comparing their representations. We employ our classifier in the Global Optimization with First-principles Energy Expressions (GOFEE) algorithm\cite{GOFEE2020,GOFEE2022},
where we use it to filter out unwanted structures from the search to improve the explorative capabilities of the algorithm. 
We show that this improves the performance of GOFEE in terms of finding the GM structure of different atomistic systems, demonstrating the usefulness of the classifier
for solving GO problems.

The paper is outlined as follows: First, the need for a reliable classifier is presented by considering a dataset of different two-dimensional pyroxene (MgSiO$_3$)$_4$ configurations.
The classifier is based on a modified adjacency matrix
combined with Voronoi tessellation, and its performance on this dataset is investigated. 
Secondly, the problem of stagnation in GO algorithms is discussed, and a structural filtering method based on the classifier is proposed.
Third, the GOFEE algorithm is explained, and the filtering method is incorporated into its sampling protocol. 
Fourth, the filtering method is used to find the global
energy minimum structure of a two-dimensional pyroxene test system.
Finally, we use the method to solve three-dimensional global optimization problems, namely finding the global energy minimum structure of olivine
(Mg$_2$SiO$_4$)$_4$, Au$_{12}$ and Lennard-Jones LJ$_{55}$ systems. We also demonstrate the usefulness of the method by solving the LJ$_{75}$ system using
a modified GOFEE algorithm.

The methods presented in this paper have been implemented in the Atomistic Global Optimization X (AGOX)\cite{AGOX2022} Python code,
which is available at \href{https://gitlab.com/agox/agox}{https://gitlab.com/agox/agox}.

\section{Methods}
\label{sec:methods}
In the first part of this section, we introduce methods to construct a classifier that can distinguish between structures
from different energy basins. 
Subsequently, we discuss how this classifier can be used in a global optimization setting, where it guides the search towards
unexplored regions of the PES.

\subsection{Filtering structures from a dataset}
\label{sec:voronoi}

\begin{figure}[]
    \includegraphics{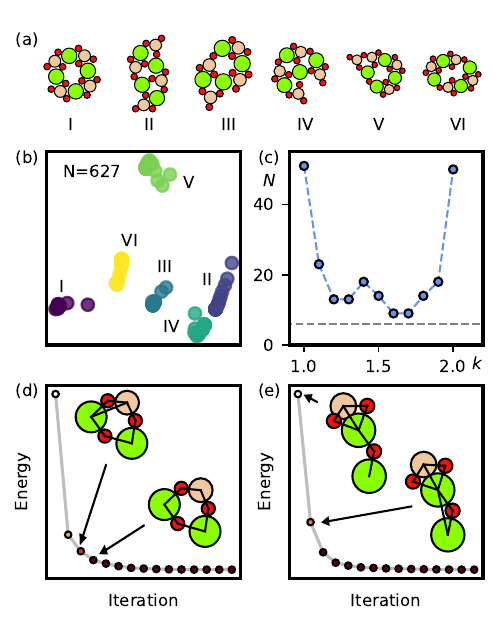}
    \caption{\textbf{(a)} The 6 distinct structures from which the dataset is constructed.
    \textbf{(b)} The structures from the database, plotted in two dimensions
    using principal component analysis. Each color
    represent different energy basins.
    \textbf{(c)}
    The result of applying the modified graph descriptor for different choices of distance cutoff $d_k$.
    The grey dashed line indicates the number of different energy basins, namely 6.
    \textbf{(d)} The energy during a local relaxation. Different colors represent different graph descriptors. The insets show a subset of atoms from the structure, where a bond between a Mg and a Si atom disappears.
    Bonds between atoms are shown as black lines.
    \textbf{(e)} Atoms from another structure, where a bond between two Mg atoms appears. Colors: Green - Mg, red - O, brown - Si.}
    \label{fig:sorting}
\end{figure}

To illustrate the task of classifying structures, we start by considering 6 distinct two-dimensional pyroxene (MgSiO$_3$)$_4$ compounds, as illustrated in Fig. \ref{fig:sorting}(a). Each of these
6 structures are randomly perturbed three times, resulting in 18 structures in total. These 18 structures
are then locally optimized in a ML energy landscape, and the trajectory from each structure relaxation is stored in a database,
which results in $N=627$ configurations. The ML energy landscape is
the Gaussian Process Regression model used in Sec. \ref{sec:2d_application}.  We calculate the Fingerprint descriptor\cite{Fingerprint2010} of these structures and plot
them in a two-dimensional space described by the first two components of a principal component analysis (PCA), which is shown in Fig. \ref{fig:sorting}(b).
See Sec. \ref{sec:fingerprint_appendix} for more details about this descriptor.
The colors represent which of the 6 energy basins each structure belongs to. We note that several of these structures have very similar descriptors,
which stems from the fact that a lot of structures are very similar near the end of a local relaxation.

Below we introduce methods that can be used to classify structures from such a dataset. These methods are based on a
simple graph representation of structures, which can be used to obtain a description of a given configuration.

\subsubsection{Modified adjacency matrix classifier}
We start by describing a classifier that is based on spectral graph theory\cite{wilson_study_2008,wills_metrics_2020}.
Here, a structure is described by an adjacency matrix, where the nodes correspond to the atoms, and the edges correspond to bonds between them. The
adjacency matrix $A^k$ is defined as

\begin{align}
    A_{ij}^k = \begin{cases}
        0, & \text{if } i=j,\\
        1 , & \text{if } d_{ij} \leq d_k,\\
        0, & \text{otherwise},
    \end{cases}
\end{align}

where $d_{ij}$ is the Euclidean distance between atoms $i$ and $j$, and 
$d_k=k \cdot (r_{{\rm cov},i} + r_{{\rm cov},j})$ is a cutoff distance, where $r_{{\rm cov},i}$ is the covalent radius of atom $i$, and $k$ is a hyperparameter.
By including the covalent radii of the atoms in the cutoff distance, the method is able to characterize bonds between different types of atoms.
If two atoms are separated by a distance which is less than the specified cutoff distance,
a value of 1 is added to the adjacency matrix, indicating that the two atoms are bound to each other.

\begin{figure}[]
    \includegraphics{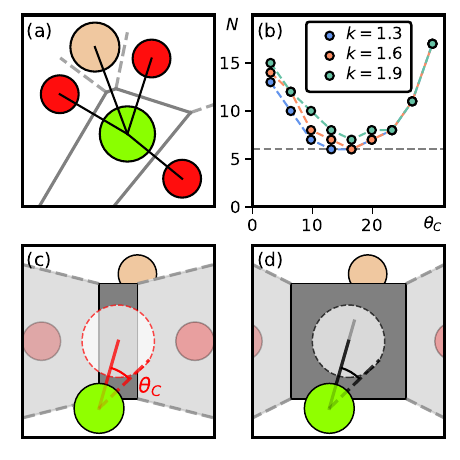}
    \caption{\textbf{(a)} A zoom in on a Mg atom. The grey lines
    indicate Voronoi faces between the atoms, and black lines indicate bonds, determined using $B^k$. 
    \textbf{(b)} The number of unique descriptors in the database as a function of the cutoff angle $\theta_C$. 
    \textbf{(c)}
    A circle centered on the possible bond between the atoms intersects with other Voronoi faces,
    meaning that the bond between the Mg and the Si atoms is removed.
    \textbf{(d)}
    A case where the circle can fit on the Voronoi face because the oxygen atoms have been moved further away from the Mg atom.}
    \label{fig:voronoi}
\end{figure}

A problem with the adjacency matrix $A^k$ is that it does
not take the atomic type of the individual atoms into account. For example, there is no way to distinguish 
between an A-A-B and an A-B-A configuration, where A and B are different atomic species. 
To solve this, we introduced a modified version of the adjacency matrix, $B^k$,

\begin{align}
    B_{ij}^k = \begin{cases}
        Z_i, & \text{if } i=j,\\
        1 , & \text{if } d_{ij} \leq d_k ,\\
        0, & \text{otherwise},
    \end{cases}
\end{align}
where $Z_i$ is the atomic number of atom $i$. 

We note that the matrix $B^k$ is
not invariant with respect to swapping indices of the atoms. Since $B^k$ is real and symmetric, its spectrum of eigenvalues
only consists of real numbers. This can be used to define a descriptor $G$ of a structure, consisting of $N_A$ atoms, as

\begin{align}
    G = [\lambda_1, \lambda_2,\dots,\lambda_{N_A}],   
\end{align}

where $\lambda_1 \leq \lambda_2 \leq \dots \leq \lambda_{N_A}$ are the $sorted$ eigenvalues of $B^k$. To avoid problems with numerical precisions, 
the eigenvalues are rounded to three decimal places. This descriptor is only based on interatomic distances, meaning that it is invariant with respect
to translations and rotations of a given structure, and it is invariant with 
respect to the indexation of the atoms. Using this descriptor, we consider two
configurations to be the same if they have identical spectra of eigenvalues, and this will be used to classify structures.

We note that for a structure containing $N_A$ atoms, the descriptor $G$ will contain only $N_A$ eigenvalues, which is less than the
3$N_A$-6 degrees of freedom of the system. This means that the descriptor is not guaranteed to be unique for a given structural configuration.
While this is not ideal when extracting unique structures from a dataset, 
we have not seen any problems due to this in practice, and
for the GO problems considered later on in this paper,
this descriptor works well in terms of finding the GM structure.

Having designed a simple classifier, we can apply it to the dataset using different scaling parameters, $k$, for the cutoff distance. The results are shown in Fig. \ref{fig:sorting}(c). 
It can be seen that the classifier is not able to obtain only one structure from each energy basin, and that the quality of the classifier depends heavily on the value of $k$.
This can be explained by the discontinuity
of the modified adjacency matrix $B^k$. Small changes to the position of atoms, that are separated by a distance close to the cutoff distance, can change the matrix $B^k$, and therefore also the descriptor $G$.
This is illustrated in Fig. \ref{fig:sorting}(d) and (e), where bonds between atoms appear or disappear during the local relaxation of structures. Thus, basing the classification of structures only on the distance between atoms is not a strong enough criterion if one needs to 
extract exactly one structure from each energy basin.

\subsubsection{Voronoi tessellation}
In order to get rid of the problem with atoms that are separated by a distance close to the cutoff distance,
we need a more robust way of describing bonds between atoms.
Looking at the structures like those in Fig. \ref{fig:sorting}(d) and (e), we often observe that a major problem
for the modified adjacency matrix is to determine the existence of bonds between Mg-Mg, Mg-Si, or Si-Si
pairs that are separated by oxygen atoms. To tackle this problem, we want to divide space around each atom into distinct regions
and quantify how much two atoms actually "see" each other. 

To do this,
we proceed by employing a Voronoi tessellation method together with the modified adjacency matrix.
This method allows for environment aware bond criteria by considering the local environment around each atom.
It works by dividing space into regions, where each region consists of all points that are closest to some
prespecified points, which in this case are the positions of the atoms. The regions are separated by planes that are placed midway between atoms. These planes are referred
to as Voronoi faces, and they can be used to determine if two atoms are bound to each other.
In this work, we have formulated the following criteria for a bond between atoms:

\begin{enumerate}
    \item The atoms share a Voronoi face.
    \item A straight line between the two atoms passes through this Voronoi face.
    \item The Voronoi face must be large enough to enclose a circle, centered on the line through the face, that spans a certain solid angle.
\label{eq:voronoi_criteria}
\end{enumerate}

Figure \ref{fig:voronoi}(a) illustrates the Voronoi faces for a Mg atom and the atoms around it as grey lines. 
The Mg atom is bound to a silicon atom and three oxygen atoms,
indicated by the black lines. The Mg and the Si atoms only share a small Voronoi face, which reflects that
these atoms barely "see" each other, since they are screened from each other by the neighboring oxygen atoms.
This is the reason for introducing the third criterion. By placing a circle on the Voronoi face, centered on the potential bond, 
we can control how large a Voronoi face must be in order for a bond to be valid.

The size of the circle is determined by a hyperparameter $\theta_C$, which we refer to as a cutoff angle. 
By introducing the hyperparameter $\theta_C$, we are now able to tune the classifier to be more or less sensitive to changes in the structure. 
Figures \ref{fig:voronoi}(c) and (d) show two examples of 
how the cutoff angle is used to determine whether a bond is invalid or valid, respectively.
The result of this criterion, using different values of $\theta_C$ for three choices of $k$, is shown in Fig. \ref{fig:voronoi}(b).
The number of structures in the dataset is greatly reduced, and the classifier is much less sensitive to the choice of scaling parameter $k$ for the cutoff distance $d_k$.
Using this method, we are able to reduce the number of structures down to 6 for certain values of $k$ and $\theta_C$,
meaning that we have a more reliable way
of distinguishing between structures from different energy basins. 

In AGOX, the Voronoi tessellation method has been implemented in a simple and computationally fast manner, where the Voronoi faces are added iteratively between atoms that
are bound to each other according to the modified adjacency matrix $B^k$. If two atoms do not obey the three criteria above,
the bond between them is removed from $B^k$. For computational convenience and speed, we only sample $N_p=8$ points on the circumference of the circle and check that they do not
intersect other Voronoi faces. These points are sampled uniformly on the circle in a deterministic manner,
such that the method always returns the same descriptor for a given structure.

For the atomic systems considered in this paper, the Voronoi tessellation method is the most time consuming part, compared to setting up and 
diagonalizing the adjacency matrix, when constructing a descriptor. However, both these tasks are much less computationally costly compared
to other parts of global optimization algorithms, for example DFT calculations. 

\begin{figure}[t!]
    \includegraphics{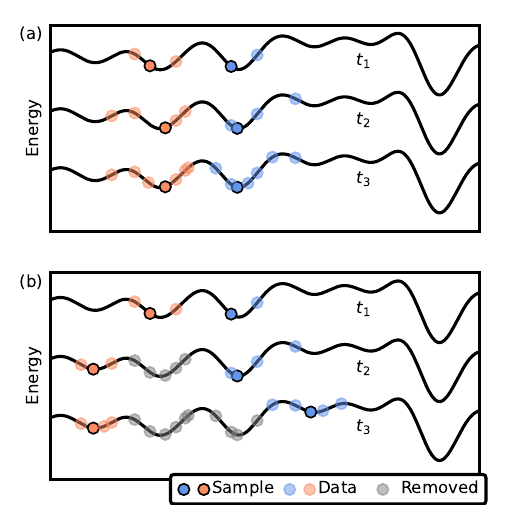}
    \caption{\textbf{(a)} A schematic example of a stagnated search at different times during a search. The search
    is stuck in a region of the PES far from the GM structure.
    \textbf{(b)} The desired effect of the filtering method. Certain structures are not allowed to enter the sample,
    which allows the search to focus on other parts of the PES.}
    \label{fig:schematic}
\end{figure}

\subsection{Filtering of structures during global optimization search}
\label{sec:go}

\begin{figure}[t]
    \includegraphics{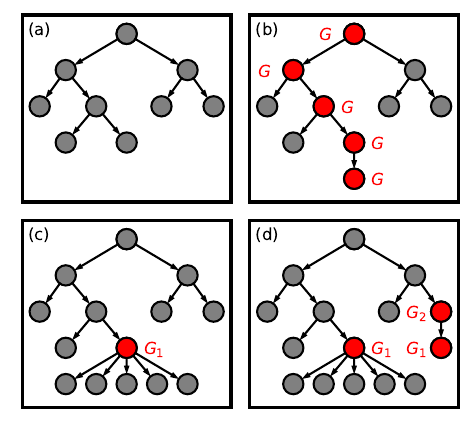}
    \caption{This figure schematically shows different situations that should be avoided during a search.
    The grey nodes correspond to structures,
    and the arrows indicate the relationship between them. In \textbf{(a)}, no structures have
    been filtered out yet. 
    The other figures show unwanted situations, namely \textbf{(b)} structures with the same descriptor get created over and over again, 
    \textbf{(c)} the same structure produces too many new structures,
    \textbf{(d)} a structure produces a new configuration with an already removed descriptor.}
    \label{fig:tree}
\end{figure}

Some GO algorithms rely on sampling a subset of known low-energy structures to produce new structures,
which introduces the risk of stagnation, because the search keeps exploiting known regions of the PES.
This is illustrated in Fig. \ref{fig:schematic}(a). Here, the two sample members are always picked from the 
same two energy basins, and the search is prone to stagnate.
To avoid this problem, a GO algorithm requires a
reliable way to distinguish between structures and filter away any unwanted subset of these,
which is illustrated in Fig. \ref{fig:schematic}(b).
To achieve this, we employ the classifier introduced in Sec. \ref{sec:voronoi}.
By tracking the descriptors of all structures that are found during the search,
the algorithm is able to make a 'family tree'-like history of all the structures and use it to filter out unwanted structures to avoid certain
scenarios to occur.

Figure \ref{fig:tree} shows schematic examples of the history of a search and such unwanted scenarios.
The grey nodes correspond to structures, and the arrows
indicate which structure was modified to generate another structure, establishing a $parent$-$child$ relation between structures. In Fig. \ref{fig:tree}(a), the search is doing fine in terms of exploring
different structures. 

Figure \ref{fig:tree}(b) shows a situation where a chain of structures with the same descriptor $G$ has been produced. 
This means that the search makes new structures that are very similar to the parent structure, but which has a slightly lower energy. This 
situation can be seen in Fig. \ref{fig:schematic}(a), where the same two local minima are being refined over and over again.
This is an unwanted behavior of the sampler, since this causes the algorithm to focus too much on few parts of the PES. Therefore,
we want to filter away structures with the descriptor $G$ to avoid this situation.

Figure \ref{fig:tree}(c) shows another situation where the descriptor $G_1$ has been used to produce five new structures. Since the same structure
with description $G_1$ enters the sample multiple times, it means that all structures generated from it have a higher energy and will not be picked
as a sample member.
We therefore want to filter out structures with this specific $G_1$ descriptor from entering the sample,
since no new structures with lower energy can be produced from them.

Finally, in Fig. \ref{fig:tree}(d), a situation is depicted where a structure with the descriptor $G_2$ has been used to produce a structure with the descriptor $G_1$. 
This indicates that the structure with descriptor $G_2$ is close to the structures with descriptor $G_1$ in the configurational space,
meaning that the search is still focusing on a region of the PES that has already been exploited. 
Since we have removed structures with the descriptor $G_1$, we also want to filter out structures with the descriptor $G_2$.

To avoid these unwanted situations from Fig. \ref{fig:tree} to occur, we require that structures with a descriptor $G$ do not produce more than $N_{\rm max}$ new structures,
and that they do not produce new structures with already removed descriptors. By doing so, we allow the algorithm to leave a region of the PES
that it has exploited, and focus on new, unknown regions, which is illustrated in Fig. \ref{fig:schematic}(b). This filtering method will be referred to as the Family Tree Filtering (FTF) method.

\section{The GOFEE algorithm}

Now that we have identified scenarios we want to avoid during a search,
we can test the FTF method in a GO algorithm. 
In this work, we focus on the Global Optimization with First-principles Energy Expressions (GOFEE) algorithm.
GOFEE is an iterative search method that combines elements from EAs with ML methods. In each iteration,
a subset of previously found structures are sampled. These sampled structures, also referred to
as $parents$, are modified to create $N_C$ new structures. The new structures are locally optimized in an uncertainty-aware machine learned energy model,
and the structure with lowest confidence bound (= predicted energy minus some degree of predicted uncertainty), according to this ML model, is evaluated in the target potential, i.e., DFT or LJ. 
More details about the GOFEE algorithm can be found in the Appendix, Sec. \ref{sec:gofee_appendix}.

The sampling of known structures to obtain the parent structures is based on the k-means algorithm. This sampling method was introduced in Ref. \onlinecite{merte_2022}, where it was shown to increase
the performance of the GOFEE algorithm. In this method, all evaluated structures with an energy $E>E_{\rm min}+5\ \rm eV$ are removed
from the pool of potential sample structures, where $E_{\rm min}$ is the lowest energy found so far. The remaining structures are clustered into $N_k$ clusters, where $N_k$ is a predefined hyperparameter
equal to the desired number of sampled structures. 
Finally, the sample of parent structures is constructed by extracting the lowest energy structure from each of the $N_k$ clusters.
A schematic one-dimensional example of this is shown in Fig. \ref{fig:schematic}(a) 
with $N_k=2$ and at different times of the search. When the FTF method is employed in GOFEE, it will be used to filter out unwanted structures
before the k-means sampling takes place, such that the sample keeps evolving during the search as illustrated in Fig. \ref{fig:schematic}(b).

We resort to success curves to compare the performance of the GOFEE algorithm with and without the FTF method. To construct a
success curve, a GOFEE search is started 100 times with the same settings but different initial atomistic configurations.
For each of these 100 searches, the initial configurations were constructed by randomly placing the atoms in a large confinement cell\cite{AGOX2022}. 
The success curve $s(x)$ then measures the proportion of these individual searches that have found a structure with the same 
descriptor as the GM structure after $x$ number of single-point calculations. We will compare success curves from different searches
that either uses the FTF method or not. Not using the FTF method corresponds to setting $N_{\rm max}=\infty$, which we will use as 
the notation for such searches.

\section{Application to a 2-dimensional test system}
\label{sec:2d_application}

\begin{figure}[hb!]
    \includegraphics{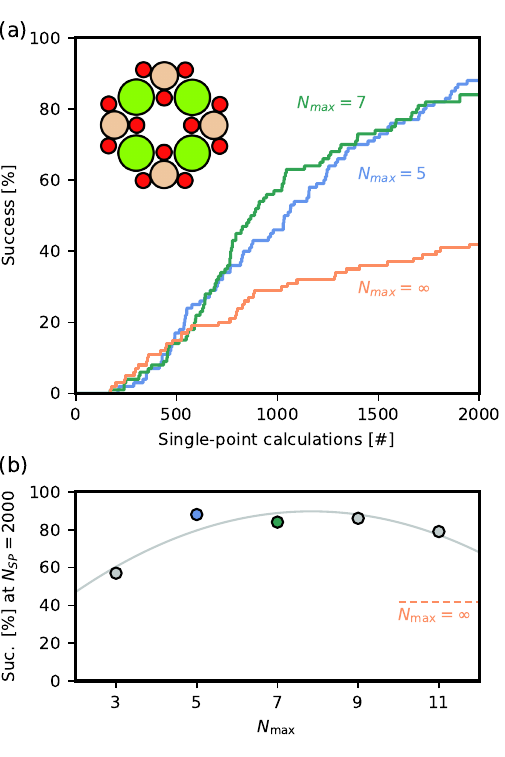}
    \caption{\textbf{(a)} The success curves of the GOFEE algorithm with the FTF method when searching for the GM structure of the 2-dimensional pyroxene system.
    \textbf{(b)} The final success after 2000 target evaluations for different choices of $N_{\rm max}$. The grey line is plotted to guide the eye.
    }
    \label{fig:succes_pyroxene}
\end{figure}

\begin{figure}[b]
    \includegraphics{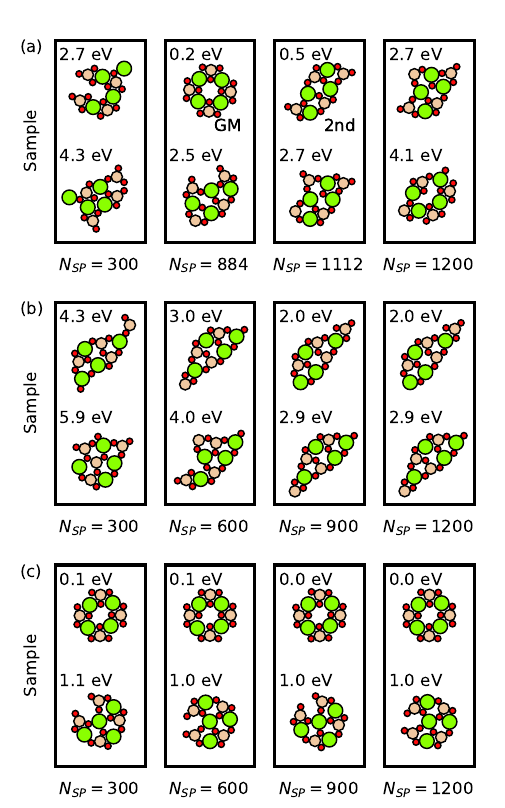}
    \caption{The evolution of the sample of \textbf{(a)} a successful search using $N_{\rm max}=5$, \textbf{(b)} an unsuccessful and \textbf{(c)} a successful GOFEE search with $N_{\rm max}=\infty$
    after $N_{SP}$ number of target evaluations. 
    The number above each structure indicates its energy relative to that of the GM structure. The GM configuration from \textbf{(a)} was not fully optimized before the FTF method filtered it away,
    meaning that a relaxation of the remaining energy of 0.2 eV was not obtained.}
    \label{fig:gofee_samples}
\end{figure}

This section shows the results of introducing the FTF method
when searching for the GM structure of the 2-dimensional pyroxene (MgSiO$_3$)$_4$ system. The target potential was a pretrained Gaussian Process Regression\cite{GPR2005} (GPR) model, trained on 314 pyroxene structures. 
A small sample size $N_k=2$ was used in order to visualize the evolution of the sample during searches. The success curves with and without
the filtering can be seen in Fig. \ref{fig:succes_pyroxene}(a). The figure shows that employing the FTF method 
increased the performance of GOFEE in terms of finding the GM structure.

Figure \ref{fig:succes_pyroxene}(b) shows the final success after 2000 target evaluations for different choices of $N_{\rm max}$. It can be
seen that the performance of GOFEE using the FTF method greatly depends on the choice of $N_{\rm max}$. If $N_{\rm max}$ is too low, e.g. 3, the final
success is lower compared to larger values of $N_{\rm max}$, but it is still better than not using the filtering at all. This shows that
filtering is indeed useful for the GOFEE algorithm.

The evolution of the sample from different searches can be compared to investigate the behavior of the sample members.
In Fig. \ref{fig:gofee_samples}(a),
the evolution of the sample in one of the successful GOFEE searches, which used the FTF method, is shown. The GM structure was found
after 884 target evaluation, while the second best 
configuration was found after 1112 target evaluations. This demonstrates that the search is indeed capable of exploring various regions of the PES
during a single search when employing the FTF method, which is the desired outcome of adding the method to the GOFEE algorithm.

This can be compared to samples from GOFEE searches that did not employ the FTF method. Figure \ref{fig:gofee_samples}(b) shows the evolution of the sample of an unsuccessful GOFEE search.
Clearly, up to iteration 900, the sample consisted of progressively lower energy structures,
but from then on the sample did not evolve, meaning that the search was stuck in some region of the PES. Fig. \ref{fig:gofee_samples}(c) shows
the sample from a successful GOFEE search. The GM structure was obtained in less than 300 iterations, 
but the sample did not change at all during the remaining part of the search, except
for some energy reduction of the already found structures. Even though this search was successful, it does not find the
second best structure, shown in Fig. \ref{fig:gofee_samples}(a), indicating that the search could not escape these low-energy regions of the PES. 

\section{Further applications}

Now that we have demonstrated the effect of adding the filtering method to GOFEE, we will apply it to solve
more interesting problems. In this section, GOFEE will be used to find the GM structure of a 3-dimensional olivine compound and a Au$_{12}$ configuration.
The target potential will be DFT using the Perdew-Becke-Ernzerhof (PBE)\cite{PBE1996} exchange-correlation functional and a LCAO\cite{lcao2009} basis using the GPAW\cite{GPAW2010} module.
In both cases, the structures were placed in a $20\times20\times20$ \AA$^3$ box. A sample size of $N_k=5$ was used in both cases.
To further show the usefulness of the FTF method in the GOFEE algorithm, we employ the computationally cheap Lennard-Jones potential to search for larger compunds. 
The GOFEE algorithm is used to find the GM structure of the LJ$_{55}$ systems. After that, we turn off the ML part of GOFEE in order to solve an even larger problem, namely the
LJ$_{75}$ system.

\subsection{Olivine compound}

In this section, GOFEE was used to solve the problem of finding the GM structure of a 3-dimensional 
olivine (Mg$_2$SiO$_4$)$_4$ compound\cite{slavensky_generating_2023}. Nanosized silicate compounds
are found to be abundant in the Interstellar Medium\cite{li_ultrasmall_2001}, and several properties of these systems have been computationally
studied\cite{tang_machine_2023,escatllar_structure_2019,goumans_hydrogen_2011,marinoso_guiu_does_2021,woodley_structure_2009,zeegers_predicting_2023}.

\begin{figure}
    \includegraphics{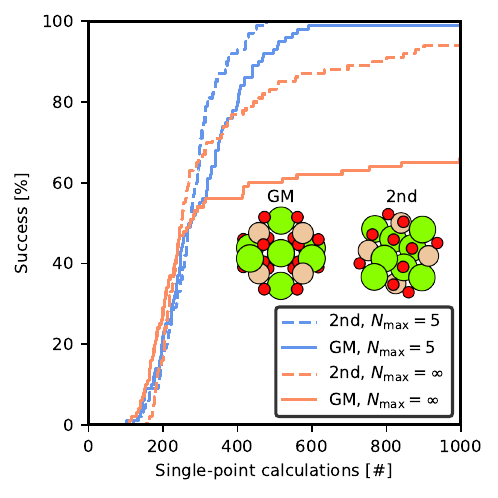}
    \caption{The success curves for finding both the GM structure and the second lowest energy structure
    during searches of the 3-dimensional olivine (Mg$_2$SiO$_4$)$_4$ system.
    For both structures, using the filtering method before sampling improved the performance
    of finding both of them.}
    \label{fig:succes_olivine}
\end{figure}

Figure \ref{fig:succes_olivine} shows the success curves for finding both the GM structure and the second-lowest energy structure.
By using the filtering method before the sampling, 
nearly all searches were able to find both the GM and the second-lowest energy structure during the same search.
It is clear that some of the searches, which did not employ the filtering method, could also find the GM, but in 1/3 of the searches
the GM was not found at all. On the other hand, it seems that almost all of them eventually find the
second best structure, but it happened much later compared to the searches that used the filtering process.
This could indicate that these searches spent too many resources exploring regions of the PES that 
were not relevant for finding the GM structure. 

\subsection{Small gold configuration}

\begin{figure}[t]
    \includegraphics{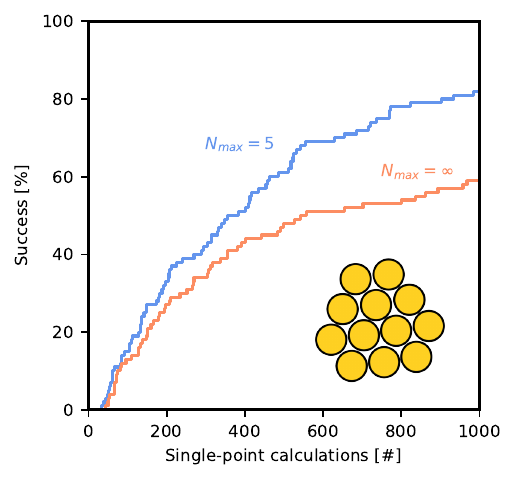}
    \caption{Succes curves for finding the GM of Au$_{12}$, shown in the inset, using the GOFEE algorithm with and without
    the FTF method.}
    \label{fig:succes_au}
\end{figure}

GOFEE was used to solve the global optimization problem of finding the global energy minimum structure of the Au$_{12}$ cluster. 
Small gold systems have previously been investigated\cite{fa_bulk_2005,xing_structural_2006,chaves_evolution_2017}.
It has been shown that the GM structure of the smaller gold clusters tend to be planar, while low-energy three-dimensional configurations exist at the same time.
On the other hand, the GM configurations of the larger compounds often appear to be three-dimensional.
This fact potentially increases the complexity of the problem, since the GOFEE algorithm might get stuck searching for 3-dimensional configurations while the GM actually is planar, and vice versa.

The planar GM structure for this global optimization problem is shown in the inset in Fig. \ref{fig:succes_au}. We allow the search to be fully three-dimensional in order to increase the difficulty of the problem.
Fig. \ref{fig:succes_au} evidences that the filtering method improves the performance of the GOFEE algorithm.

\subsection{LJ$_{55}$}
\begin{figure}[t]
    \includegraphics{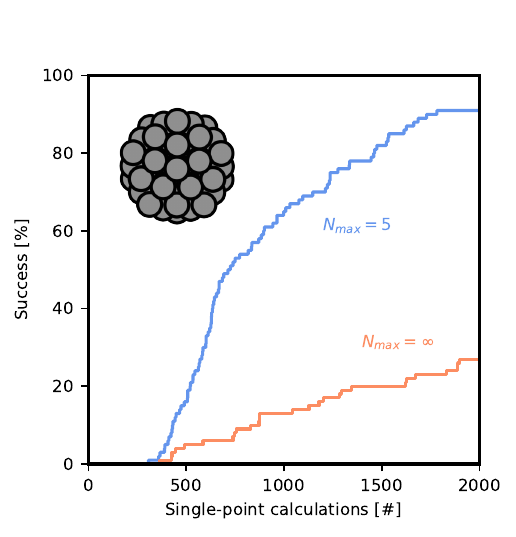}
    \caption{The succes curves for finding the GM of the LJ$_{55}$ system with and without
    the FTF method.}
    \label{fig:succes_lj55}
\end{figure}

To further examine the performance of the FTF method, we wanted to 
test it on a larger system. To do this we used the Lennard-Jones potential,
since the energy evaluations are fast to perform compared to DFT, and
because the GM structures of various LJ systems are well-known\cite{wales_global_1997}.
We therefore applied the method to the
LJ$_{55}$ system. This system has three highly symmetric minima,
making it a potentially difficult problem to solve.

The success curves for different searches can be seen in
Fig. \ref{fig:succes_lj55}. The figure demonstrates that the GOFEE algorithm greatly benefits
from using the filtering method, since the problem can be solved almost every 
time within 2000 single-point evaluations.

\subsection{LJ$_{75}$}
\begin{figure}[t]
    \includegraphics{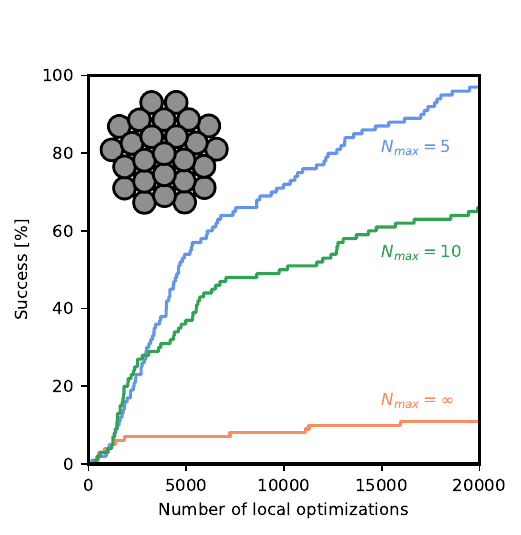}
    \caption{Different succes curves when using a modified GOFEE algorithm with and without the FTF method to find the GM of the LJ$_{75}$ structure.}
    \label{fig:succes_lj75}
\end{figure}

We now turn our attention to the LJ$_{75}$ system\cite{wales_global_1997}. For a system of this size, the configurational space
is much larger compared to the LJ$_{55}$ system, meaning that a large number of energy evaluations are
required to solve the problem. At the same time, such evaluations using the Lennard-Jones formula are quite fast,
meaning that the machine learning part of GOFEE quickly becomes the time-dominating factor. To combat this problem,
we modified the GOFEE algorithm such that no ML is used during the search. All structures are locally optimized
directly in the LJ potential instead, making it possible to perform longer global optimization searches. More details about
the modified GOFEE algorithm can be found in the Appendix, sec. \ref{sec:modified_gofee_appendix}.

The succes curves for solving this problem can be seen in Fig. \ref{fig:succes_lj75}. Without the FTF method,
the problem is really difficult to solve for the modified GOFEE algorithm, indicating that the search potentially
becomes stuck in unwanted regions of the PES. Using the FTF method
greatly improves the chance of finding the GM configuration, demonstrating that the method increases
the performance of the GO algorithm.

\section{Conclusion}
We have proposed a method for classifying structures from different energy basins in a dataset.
This method employed spectral graph theory and the Voronoi tessellation method to make it robust to
small variations to the structures, and we have shown its capability of distinguishing structures
from different energy minima. The classifier was introduced to the GOFEE algorithm as a filtering mechanism
for the sampling of known structures. By doing so, the algorithm was able to abandon regions of the PES that it had exploited
and thereby avoid stagnation.
We have shown that the filtering method greatly improves the performance
of the GOFEE algorithm when searching for an olivine (Mg$_2$SiO$_4$)$_4$, a Au$_{12}$, and a LJ$_{55}$ system, and that
a modified GOFEE algorithm benefits from using the filtering method when solving the LJ$_{75}$ system.

\section{Acknowledgements}
This work has been supported by VILLUM FONDEN through Investigator grant, project no. 16562,
and by the Danish National Research Foundation through the Center of Excellence “InterCat” 
(Grant agreement no: DNRF150).

\section*{Data availability}
The data that support the findings of this study and the corresponding AGOX scripts are available
at \href{https://gitlab.com/aslavensky/classifier_paper}{https://gitlab.com/aslavensky/classifier\_paper}. The AGOX code can be obtained from \href{https://gitlab.com/agox/agox}{https://gitlab.com/agox/agox}. 
The documentation for the AGOX module can be found at \href{https://agox.gitlab.io/agox}{https://agox.gitlab.io/agox}.

\section*{References}
\bibliography{bib}

\cleardoublepage
\section{Appendix}
\setcounter{equation}{0}
\renewcommand{\theequation}{{\thesubsection}\arabic{equation}}
\subsection{Fingerprint descriptor}
\label{sec:fingerprint_appendix}
To train a machine learning model, a translational and rotational invariant descriptor of atomic systems is required. In GOFEE,
the Fingerprint descriptor\cite{Fingerprint2010} by Valle and Oganov is used.
Here, the radial part of the descriptor is evaluated as
\begin{align*}
    F_{AB}(r) &\propto \begin{cases}
        \sum_{i,j} \frac{1}{r_{ij}^2} \text{exp} \left(- \frac{(r-r_{ij})^2}{2 l_r^2} \right)&, \quad r<R_r\\
        0 &, \quad r \geq R_r
    \end{cases}\\
\end{align*}

and the angular part is calculated as

\begin{align*}
    F_{ABC}(\theta) &\propto \sum_{i,j,k} f_c(r_{ij}) f_c(r_{ik}) \text{exp} \left(- \frac{(\theta - \theta_{ikj})^2}{2l_\theta^2 } \right),
\end{align*}
where $A$, $B$, and $C$ are atomic types, $R_r=6$ Å is a radial cutoff distance, $l_r=0.2$ Å, $l_\theta=0.2$ rad, and $f_c$ is a cutoff function.
For a given structure, all two-body (radial) and three-body (angular) combinations of the atomic types are calculated and binned into 30 bins.
These bins are concatenated into a single feature vector that describes the whole atomistic system.

\subsection{The GOFEE algorithm}
\label{sec:gofee_appendix}
In this work, we focus on the GOFEE algorithm\cite{GOFEE2020,GOFEE2022}. GOFEE combines elements from EAs and machine learning (ML) to reduce the
required number of target potential evaluations required to find the GM structure. To avoid spending a lot of computational resources on doing local optimizations
of new structures in the target potential, which could become time-consuming when using DFT, 
GOFEE employs a ML surrogate model for the structure relaxation. The surrogate model, which is 
trained on-the-fly, is constructed using Gaussian Process Regression (GPR)\cite{GPR2005}. Given a database of training data $\textbf{X}$, which is a matrix
consisting of the Fingerprint representations\cite{Fingerprint2010} of the structures, and a vector of their corresponding energies $\textbf{E}$, the GPR model can predict
the energy of a new structure $\textbf{x}_*$, $E_{GPR}(\textbf{x}_*)$, and the model's uncertainty of the predicted energy, 
$\sigma_{GPR}(\textbf{x}_*)$, as

\begin{align}
    E_{GPR}(\textbf{x}_*) &= K(\textbf{x}_*,\textbf{X}) \textbf{C} (\textbf{E}- \bm{\mu}) + \mu(\textbf{x}_*),\\
    \sigma_{GPR}(\textbf{x}_*) &= K (\textbf{x}_*,\textbf{x}_*) - K(\textbf{x}_*,\textbf{X}) \textbf{C} K(\textbf{X},\textbf{x}_*),
\end{align}
where $\bm{\mu}$ is a vector containing the prior values of each training structure, $K$ is the kernel function, and 
$\textbf{C} = [K(\textbf{X},\textbf{X}) + \sigma_n^2 \textbf{I}]^{-1}$, where $\sigma_n=10^{-2}$ eV acts as regularization.
In GOFEE, the kernel $K$ is given as 

\begin{align}
    \nonumber K(\textbf{x}_i,\textbf{x}_j)  =  
    &\theta_0(1-\beta) \text{exp}  \left( - \frac{1}{2} \left[\frac{\textbf{x}_i - \textbf{x}_j} {\lambda_1} \right]^2 \right) + \\ &\theta_0 \beta \text{exp}\left( - \frac{1}{2} \left[\frac{\textbf{x}_i - \textbf{x}_j} {\lambda_2} \right]^2 \right) 
\end{align}

where $\beta=0.01$ is the weight between the two Gaussians, $\theta_0$ is the amplitude of the kernel, and $\lambda_1>\lambda_2$ are length scales. The prior $\mu$ is given by

\begin{align}
    \mu(\textbf{x}) = \bar{E} + \frac{1}{2} \sum_{ab} \left( \frac{1 \text{Å}}{r_{ab} + 1\text{Å} - 0.7r_{CD,ab}}\right)^{12} \text{eV},
\end{align}
where $r_{ab}$ is the distance between atoms $a$ and $b$, $r_{CD,ab}$ is the sum of their covalent radii, and $\bar{E}$ is the mean energy of the training data.

Using these quantities, a lower confidence bound (LCB) energy expressions can be formulated,

\begin{align}
    E_{\rm LCB}(\textbf{x}_*) = E_{GPR}(\textbf{x}_*) - \kappa \sigma_{GPR}(\textbf{x}_*),
\end{align}
where $\kappa=2$ is a hyperparameter that controls the trade-off between exploration and exploitation of the PES.
This energy model is used in GOFEE to do local optimization of structures.
More details about the GPR model can be found in Ref. \onlinecite{GOFEE2022}.

The GOFEE algorithm starts out by making completely random structures, evaluating their energies, adding them to a database, and construct the initial surrogate model.
From then on, it iteratively samples the database, makes new structures, and improves the surrogate model in its attempt
of finding the GM structure. GOFEE can be summarized as follows:

\begin{enumerate}
    \item Sample $N_k$ parent structures from the database of previously evaluated structures.
    \item Modify these parent structures to obtain $N_C$ new structures.
    \item Locally optimize all $N_C$ new structures in the LCB landscape.
    \item Pick the structure with the lowest energy predicted by the LCB energy expression.
    \item Evaluate its energy in the target potential.
    \item Add the new structure and its energy to the database. Retrain the surrogate model.
    \item Repeat steps 1-6 until a certain number of iterations has been done.
\end{enumerate} 

\subsection{Modified GOFEE algorithm}
\label{sec:modified_gofee_appendix}
The difference between the GOFEE algorithm and the modified version of it, used to solve the LJ$_{75}$ system,
is that the modified GOFEE algorithm does not employ ML. This choice was made to investigate the performance 
of the FTF method for a large well-known problem. The modified GOFEE algorithm can be summarized as follows:
\begin{enumerate}
    \item Sample $N_k$ parent structures from the database of previously evaluated structures.
    \item Modify these parent structures to obtain $N_C$ new structures.
    \item Optimize all $N_C$ structures in the LJ potential and add them to the database.
    \item Repeat steps 1-3 until a certain number of iterations has been done.
\end{enumerate} 

\end{document}